\documentclass{appolb}
\usepackage{graphicx}
\usepackage{epstopdf}

\begin{document}
\title{Accuracy of fission dynamics within the time dependent superfluid local density approximation%
\thanks{Presented at the XXXV Mazurian Lakes Conference on Physics, Piaski, Poland, September
3-9, 2017}%
}
\author{J. Grineviciute
	\address{Faculty of Physics, Warsaw University of Technology,
		ulica Koszykowa 75, 00-662 Warsaw, POLAND \linebreak}\\ 
	P. Magierski
	\address{Faculty of Physics, Warsaw University of Technology,
		ulica Koszykowa 75, 00-662 Warsaw, POLAND\\
	Department of Physics, University of Washington, Seattle, WA 98195--1560, USA \linebreak}\\
	A. Bulgac, S. Jin
	\address{Department of Physics, University of Washington, Seattle, WA 98195--1560, USA \linebreak}\\
	I. Stetcu
\address{Theoretical Division, Los Alamos National Laboratory, Los Alamos, NM 87545, USA}
}

\maketitle
\begin{abstract}
 We investigate properties of the method based on time dependent superfluid local density approximation
 (TDSLDA) within an application to induced fission of $^{240}$Pu 
 and surrounding nuclei. Various issues related to accuracy of time evolution and the determination
 of the fission fragment properties are discussed.
\end{abstract}
\PACS{21.60.Jz, 25.85.-w, 25.70.-z}
  
\section{Introduction}
The time-dependent superfluid local density approximation (TDSLDA) is an extension of time-dependent 
Density Functional Theory (TDDFT) to superfluid systems using local pairing field. It is designed to 
describe real-time dynamics of inhomogeneous fermionic systems subjected to perturbations of 
arbitrary strength. The method is flexible and allows for applications in various quantum
systems which are defined through a suitable energy density functional. It is also microscopic
in a sense that fermionic degrees of freedom are treated and evolved explicitly 
without any additional phenomenological input. 
Various successful applications of TDSLDA include
physics of ultracold atomic gases \cite{Bu09}, low energy nuclear physics  \cite{Io11,Io16}, and 
physics of the neutron star crust \cite{Ga16}. The main advantage of TDSLDA consists of 
treating paired fermions as a dynamic field which has its own modes of excitations.  
Indeed, the correct treatment of pairing is crucial when describing nuclear reactions, such as fission, 
that are strongly influenced by pairing correlations \cite{Io16,Br05}.

In the context of nuclear reactions, the typical scenario in which TDDFT is used consists of
a system being initially in a ground state or a state characterized by a certain deformation achieved using
a constraint solution (obtained within the standard DFT). Subsequently, the nucleus is acted upon 
by a perturbation that drives it out of equilibrium. The external perturbation in the nuclear system
can be of various origins: it can be caused by photon absorption, by neutron capture, or
the perturbation can arise as an interaction between the projectile and the target nucleus. 
It has to be emphasized that the perturbation may be of arbitrary strength
since TDDFT can be applied both in the linear-response regime as well as
in the nonlinear regime. In particular, the external perturbations can be strong enough to
compete with, or even override the internal interactions that provide the structure and
stability of atomic nucleus, as in the case of induced fission process. 

The typical procedure used in the context of nuclear reactions is the following:
\begin{itemize}
\item Prepare the initial state by solving static Kohn-Sham equations for a nucleus (or nuclei if
more than one system is involved in the reaction process),  to get a
set of ground-state Kohn-Sham orbitals and orbital energies.
\item The time evolution can be obtained by applying certain external field simulating
e.g., the photon absorption, or through generating nonzero velocities of nuclei towards each other.
Then one solves the time-dependent Kohn-Sham equation from the initial time to
the desired final time. The time propagation of the orbitals determines the time-dependent densities.
\item During time evolution  one may calculate the desired observable(s) as 
functionals of densities used as building blocks of the energy density functional.
It implies that TDDFT is particularly well suited to calculate one-body observables. 
\end{itemize}

Hence, it is clear that in particular for calculations of induced fission the stability of the evolution 
of nuclear system is crucial as it influences the nascent fragment properties. In this paper, we discuss 
issues pertaining to accuracy of induced fission of selected Pu isotopes within TDSLDA employing 
Skyrme SLy4 nuclear energy density functional (NEDF).

\section{Time step and pairing coupling constant}

The set of equations originated from TDSLDA has the following form: 
\begin{eqnarray} \label{tdslda}
i\hbar\frac{\partial}{\partial t} 
\left  ( \begin{array} {c}
  U_{\mu}({\bf r},t)\\  
  V_{\mu}({\bf r},t)\\ 
\end{array} \right ) =
\left ( \begin{array}{cc}
h({\bf r},t)&\Delta({\bf r},t)\\
\Delta^*({\bf r},t)&-h^* ({\bf r},t)
\end{array} \right )  
\left  ( \begin{array} {c}
  U_{\mu}({\bf r},t)\\
  V_{\mu}({\bf r},t)\\ 
\end{array} \right ),
\end{eqnarray}
where $h$ and $\Delta$ are determined by the energy density functional through the densities
which are constructed from amplitudes $U$ and $V$ (for clarity we omit here spin and nucleon indices).
Therefore, TDSLDA represents a system of coupled, nonlinear time-depended partial differential equations. 
In nuclear applications, the number of equations is of the order of hundreds of thousands (depending on the lattice size).
The time interval of the evolution of the system depends on the physical problem studied, and in the case
of induced fission it is set by the time needed for a nucleus to move from saddle point to separated fragments, 
which requires around half a million time steps. 

In the case of induced fission, the time-depended equations of TDSLDA are solved on a spatial lattice 
discretized by a lattice constant $a$. The lattice spacing
$a=1.25$ fm sets the cubic momentum cutoff $p_{c}=\hbar k_{c} \approx 860$ MeV/c, where $k_{c}= \sqrt{3} \pi /a$ \cite{Bu13}. 
For fission of $^{240}$Pu, $^{238}$Pu and $^{242}$Pu, the $(25 \ \rm{fm} )^{2} \times 50 \ \rm{fm}$
 box is used,  which is large enough to determine properties of two separated fission fragments. 

\begin{figure}[htb]
\centerline{	\includegraphics[width=\textwidth]{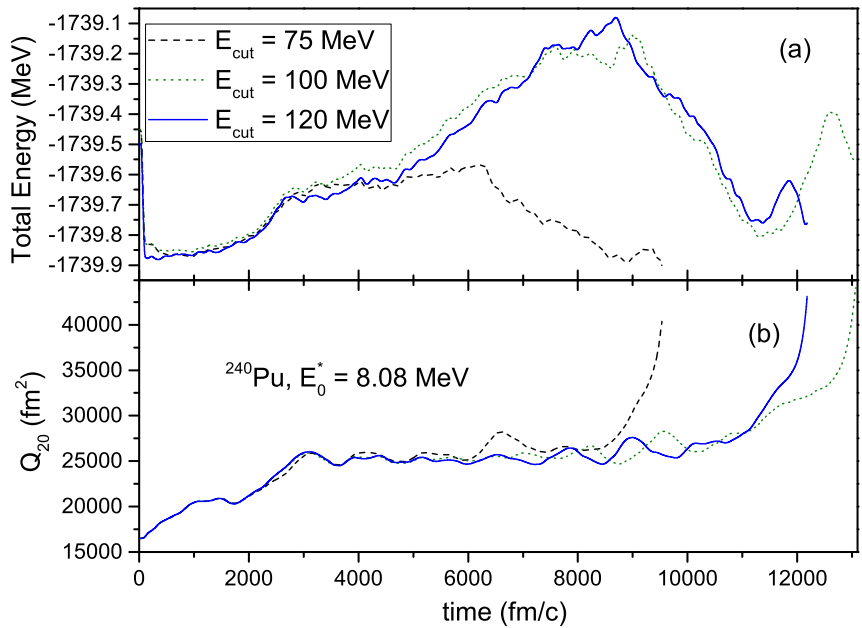}}
	\caption{(Color online) Time evolution of induced fission of $^{240}$Pu compound nucleus with pure volume pairing
		on a $18^{2} \times 40$ lattice  with a
		lattice constant of 1.25 fm at $E^{*}=8.08$ MeV
		using a spherical cutoff for 3 different cutoff energies: 
		75 MeV (black dashed line), 100 MeV (green dotted line), 120 MeV (blue solid line).
		Subfigure (a) shows total energy as a function of time, and subfigure (b) shows quadrupole Q$_{20}$ moment as a function of time.}
	\label{Figs3}
\end{figure}

The time evolution is performed  using the fifth-order Adams-Bashforth-Milne predictor-modifier-corrector method \cite{Ha12}. 
The number of evolved amplitudes $U_{\mu},V_{\mu}$  is equal to $4 N_{xyz}$, where $N_{xyz}$ is the number of lattice points
(factor $4$ originates from two spin and two isospin values). This is due to the fact
that Bogoliubov transformation, which is defined at each time step, has to fulfill completeness relations and, thus,
requires to include all states within the space defined by the lattice. Note, that this is in contrast to  the static
SLDA calculations, where the energy cutoff can be set at much lower energies. In TDSLDA, one can prove that the energy
of the system is conserved only when all states are evolved  (see Refs. \cite{Mag2016} for discussion
related to the energy cutoff in TDDFT). 
In practice, however, if only a short time evolution is required
imposing a lower energy cutoff is enough as the discrepancies occur only after a certain time interval.
It is illustrated in Fig. \ref{Figs3}, where $^{240}$Pu is evolved at 
$E^{*}=8.08$ MeV on a $(22.5 \ \rm{fm} )^{2} \times 50 \ \rm{fm}$ lattice for 3 different cutoff energies:
75 MeV (black dashed line), 100 MeV (green dotted line), 120 MeV (blue solid line). The relation between pairing coupling
constants for three energy cutoffs is set according to Refs. \cite{BuYu02}.
Clearly, the discrepancy between these three solutions is cutoff dependent. However, for short time scales 
even a relatively low energy cutoff is sufficient to obtain accurate solutions.  

In the induced fission studies, however, a long time evolution is required and,
therefore, one needs to include all the states. The pairing coupling constant needs to be renormalized through the relation \cite{BCS-BEC}
\begin{eqnarray} 
\label{eqcutoff}
\Delta \left(\textbf{r}\right)=-g_{eff} \chi \left(\textbf{r}\right) \nonumber\\
\frac{1}{g_{eff}}=\frac{1}{g} -   \frac{m_{eff} K}{4 \pi \hbar ^{2}b} .
\end{eqnarray}
where $ \chi$ is the anomalous density, $b$ is the lattice constant and $K= 2.4427496$ is a numerical factor, resulting from the expression:
\begin{equation} 
K=\frac{12}{\pi} \int_{0}^{\pi /4} d \theta \ln (1+1/ \cos ^2 \theta)
\end{equation}
as in Ref. \cite{BCS-BEC}. 

\begin{figure}[htb]
	\centerline{	\includegraphics[width=0.95\textwidth]{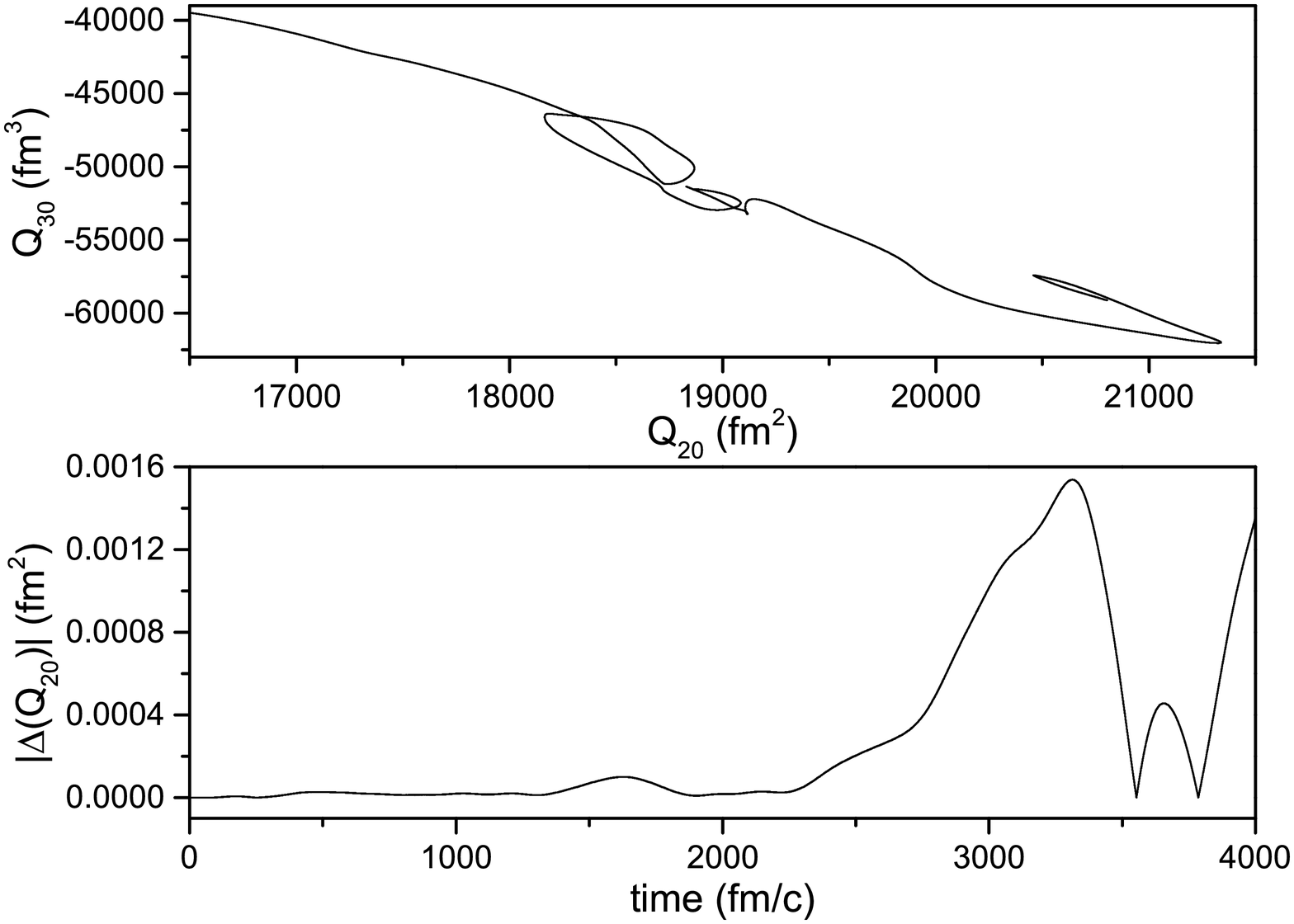}}
	\caption{(Color online) Time evolution of fissioning $^{242}$Pu up to 4000 fm/c. 
		Subfigure (a) shows octupole Q$_{30}$ moment as a function of quadrupole Q$_{20}$ moment (time step: $\Delta t=0.030$ fm/c). 
		Subfigure (b) shows the differences in the quadrupole moment for two evolutions obtained with time steps: $\Delta t=0.030$ fm/c and $\Delta t=0.024$ fm/c}
	\label{Figs1}
\end{figure}

Using the cubic cutoff renormalization and evolving all the states, the energy cutoff $E_{max} \approx 400$ MeV 
is three times larger as compared to calculations with spherical cutoff \cite{Io16} and, therefore, a time step $\Delta t \sim 1/E_{max}$ for the dynamic calculation needs to be reduced. 
Consequently, the time step $\Delta t=0.03$ fm/c ($\approx 10^{-25}$s) turns out to be sufficient to get a stable 
solutions within the required time interval. The relative error for evolutions with two different
time steps is shown in Fig. \ref{Figs1}. Namely, the time evolution of $^{242}$Pu nucleus at $E^{*}=4.90$ MeV 
for $T<4000$ fm/c  with  $\Delta t_{0} = 0.03$ fm/c and  $\Delta t_{0} = 0.024$ fm/c has been plotted. 	
Note that for this case, the total energy difference between the two trajectories with two different time steps 
is within $1$ eV!

\section{Energy conservation}

\begin{figure}[htb]
	\centerline{	\includegraphics[width=\textwidth]{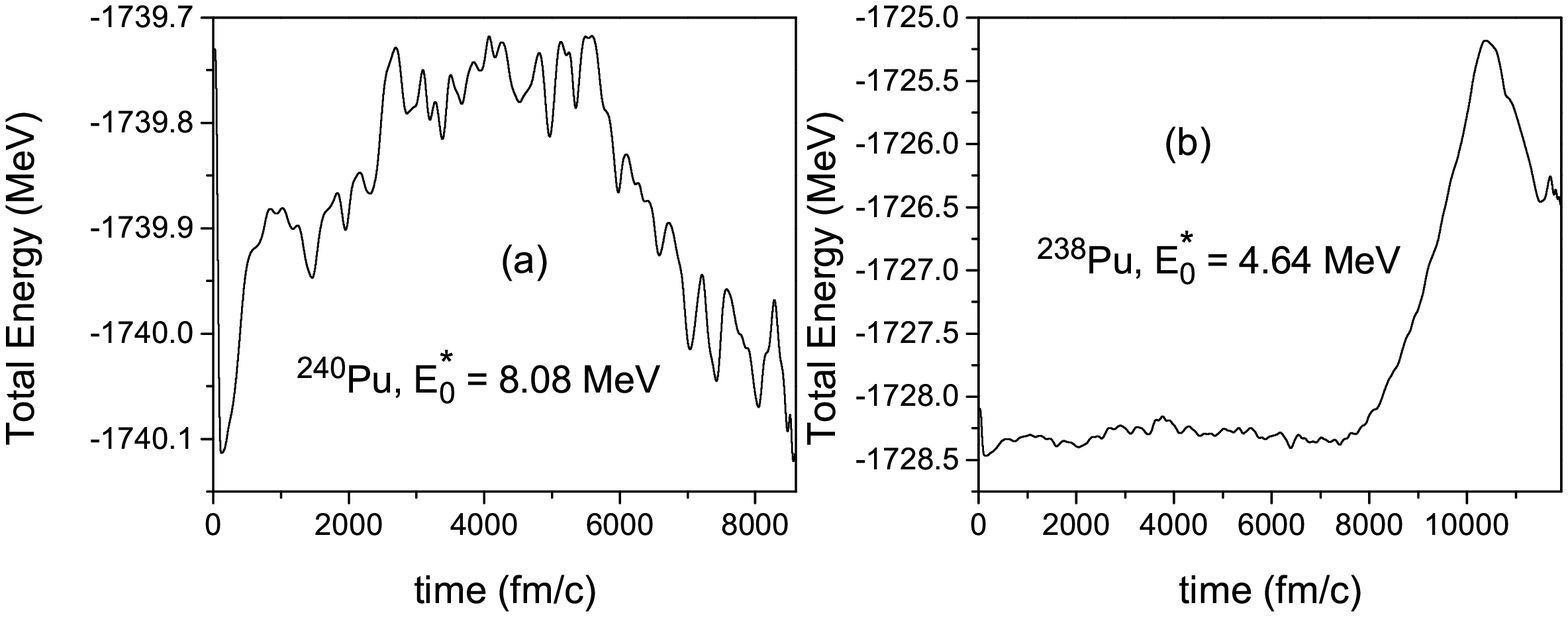}}
	\caption{ Total energy as a function of time during saddle-to-scission induced fission. Subfigure (a) shows fission of  $^{240}$Pu compound nucleus, and subfigure (b) shows $^{238}$Pu compound nucleus. Initial states of compound nuclei with a quadrupole constraint  $q_{0}=16500$  fm$^{2}$ correspond to $E_{x}=8.08$ MeV and  $E_{x}=4.64$ MeV, respectively.}
	\label{FigE}
\end{figure}

The initial state for the simulation of the fission process has been prepared by solving the static DFT equations with the constraint, which
produced reflection asymmetric shape with a quadrupole moment  $q_{0}\approx 16500$ fm$^{2}$.
It corresponds to an excited configuration slightly beyond the outer barrier ($q_{0}\approx 14000$ fm$^{2}$). 
The initial state of $^{240}$Pu nucleus is 8.08 MeV above the ground state and  corresponds to 
the neutron incident energy $E_{n}=1.54$ MeV . Analogously, the constrained static solution for 
$^{238}$Pu nucleus is $4.64$ MeV above the ground state. 
The constraint has been released adiabatically, and
the number of steps in the saddle-to-scission time evolution is about $300-500$ thousand. The average particle number in time-evolution is conserved. The precision of the total energy conservation of the system during the time evolution varies from 0.5 MeV to 3.5 MeV, which is 0.03\% to 0.2\% of the total energy, respectively (see Fig. \ref{FigE}). 

Note, that as shown in Fig. \ref{Figs1} (a) the fission path for Pu isotopes is not smooth.  At various times during evolution, the nucleus 
rearranges its structure and various internal degrees of freedom are excited 
along the fission path. When the nucleus keeps rearranging, but cannot find a fission path, the error in total energy keeps increasing and the nucleus does not fission. 


\section{TKE and excitation energies of the fragments}

Within the framework of TDSLDA, one can easily extract the total kinetic energy (TKE) and, hence, the total excitation energy (TXE) of the fragments, as well as the excitation and kinetic energies of each fragment separately. 
It is one of the main advantages of TDSLDA over those based on adiabatic assumptions or 
treating internal degrees of freedom within semiclassical approaximation (e.g., based on the Langevin-type equation).
TKE of the fragments can be extracted when fragments are well separated, i.e., at distances larger than $10$ fm.
The typical accuracy of TKE determination is about 0.5 MeV, that is 0.3 \%.

The agreement of TKE with experimental data for  $^{239}$Pu(n,f) at $E_{n} < 6$ MeV  \cite{Aki71} is within 3.5 MeV, which is 
corresponds to 2 \% error. For example: $ \rm {TKE}(E_{n}=1.54 \rm{MeV})= 173.8$ MeV (exp. 177.3 MeV), $ \rm {TKE}(E_{n}=5.63 \rm{MeV})= 176.1$ MeV (exp. 175.8 MeV). 

\section{Conclusions}

The superfluid TDDFT in the framework of TDSLDA is a perfect candidate to provide a fully microscopic description of nuclear fission and low energy nuclear reactions,
among other approaches that have been suggested over the years, with various degrees of theoretical assumptions about the character of the fission dynamics (see \cite{meth} and references therein).
As we discussed above, the method, when applied on a leadership-class supercomputers, is capable
to provide results concerning fission like TKE, TXE and provide energy sharing between fragments, i.e., data
which can be compared directly to experiment. As discussed above, the accuracy of the description
of the fission process within TDSLDA is within $0.1$ \% of the total energy and the calculated TKE
allow for the $0.2$\% precision.
These values can be gradually decreased by employing higher order integration methods and smaller lattice constants.

\section{Acknowledgements}
Authors acknowledge support of Polish National Science Centre (NCN) Grants: decision 
no. DEC-2013/08/A/ST3/00708 and UMO-2016/23/B/ ST2/01789. 
We acknowledge PRACE for awarding us access to resource Piz Daint based in Switzerland at Swiss National Supercomputing Centre (CSCS), decision No. 2016153479. We also acknowledge Interdisciplinary Centre for Mathematical and Computational Modelling (ICM) of Warsaw University for computing resources at Okeanos (grant No. GA67-14).

\end{document}